\documentclass[author]{vgtc}                          

\ifpdf
  \pdfoutput=1\relax                   
  \usepackage{graphicx}                
  \DeclareGraphicsExtensions{.pdf,.png,.jpg,.jpeg} 
\else
  \usepackage{graphicx}                
  \DeclareGraphicsExtensions{.eps}     
\fi%

\graphicspath{{figures/}{pictures/}{images/}{./}} 

\usepackage{microtype}                 
\PassOptionsToPackage{warn}{textcomp}  
\usepackage{textcomp}                  
\usepackage{mathptmx}                  
\usepackage{times}                     
\usepackage{cite}                      
\usepackage{tabu}                      
\usepackage{booktabs}                  
\usepackage[T1]{fontenc}
\usepackage[hidelinks]{hyperref}

\usepackage{enumitem,hyperref}

\onlineid{5022}

\vgtccategory{Research}

\vgtcinsertpkg

\title{Intelligent Drill-Down: Large Language Model-Driven Drill-Down Technique for Human-AI Collaborative Visual Exploration}

\author{%
        Zhijun Zheng\thanks{e-mail: Simingchen@fudan.edu.cn}\\ %
        \scriptsize Fudan University 
    \and   Tian Qiu\\ %
        \scriptsize Fudan University 
    \and   Yuheng Zhao\\ %
        \scriptsize Fudan University 
    \and   Siming Chen\\ %
        \scriptsize Fudan University 
}

\authorfooter{
  \item
  Zhijun Zheng is with School of Computer Science and Artificial Intelligence, Fudan University. Tian Qiu, Yuheng Zhao, Siming Chen are with School of Data Science, Fudan University. Siming Chen is also with Ji Hua Laboratory and Shanghai Key Laboratory of Data Science. Siming Chen is the corresponding author (simingchen@fudan.edu.cn)
}

\abstract{%
  In visual analytics, applying filters to drill-down and extract higher-value insights is a common and important data analysis method. When the drill-down space becomes excessively large, analysts may lose orientation, leading to decreased efficiency in the drill-down process. To tackle these challenges, we propose the Intelligent Drill-Down Framework, in which a large language model (LLM) facilitates the generation of visual insights, leverages user interaction data to interpret user intent, and generates appropriate drill-down paths. Our method is designed to assist users in identifying valuable drill-down paths when exploring multidimensional data, thereby reducing the cognitive burden of data interpretation and facilitating the generation of insights. Specifically, we propose a drill-down path recommendation method, in which the LLM is trained to approximate a validated greedy algorithm. Secondly, we analyze the user's intent to construct a drill-down chart. Finally, we design a branch management method. Building upon this framework, we designed a system that includes a hybrid interface providing hierarchical navigation to monitor users and manage parallel branches, a visualization panel for interactive data exploration, and an insight panel to present analytical findings and generate drill-down recommendations. We evaluated the effectiveness of our method through a demonstrative use case and a user study.
}


\usepackage{xcolor} 
\colorlet{red}{black} 

\begin{document}


\firstsection{Introduction}

\maketitle

\textcolor{red}{In the realm of visual analytics, drill-down serves as a cornerstone technique for navigating complex information spaces. It is an iterative focus refinement process, where analysts, triggered by interactive operations such as clicking visual elements, brushing data ranges, or issuing queries, navigate from a broad analytical context to specific data subdomains~\cite{tan2019drill,chaudhuri1997overview,boncz1998drill}. Beyond traditional hierarchical traversal (e.g., temporal granularity), this navigation encompasses semantic expansion based on intrinsic data relationships (e.g., from organizations to affiliated members)~\cite{kehrer2010interactive}. Distinct from static "overview and detail" paradigms, a drill-down triggers a visualization context update: by applying specific logical predicates or topological constraints, the visual representation transitions from macro-aggregation to micro-specification, facilitating the extraction of high-value in-depth insights ~\cite{elmqvist2009hierarchical,stolte2002polaris}}.
However, drill-down operations require considerable user effort, which compromises their overall effectiveness. This inefficiency is attributable to several underlying challenges: Information overload and missed insights from decision-space explosion~\cite{lee2019avoiding,nemeth2021deep}, low-novelty drill-down paths and wasted time and effort~\cite{vartak2015seedb,zhang2007effective}, intent-divergent visualization resulting from dependence on predefined rules and graphical interactions~\cite{gao2015datatone,tan2019drill} and inadequate context management during drill-down, which degrades readability~\cite{derthick2000data,lam2008framework}.


To solve these challenges and reduce the cognitive burden on users, researchers have explored more intelligent drill-down techniques~\cite{khosravi2021intelligent}. One line of work emphasizes guiding users toward meaningful exploration paths~\cite{demiralp2017foresight,vartak2014seedb}. For instance, Smart Drill-Down~\cite{joglekar2017interactive} employs a rule-based strategy to automatically highlight potentially high-value directions, helping users filter out irrelevant branches during exploration. Building on this idea, other studies~\cite{zhao2024leva,tian2024chartgpt,wongsuphasawat2017voyager} have proposed human-computer collaboration frameworks that go beyond simple rules. A representative example is LightVA~\cite{zhao2024lightva}, which integrates interactive visual analytics with adaptive system feedback, with the aim of better supporting sensemaking in complex data environments. However, manually specified rules still require users to make their own selections and decisions, placing high demands on the manner in which they determine their course of action. The application of LLMs in drill-down remains a significant yet challenging area of unexplored research.

Our research primarily investigates how LLMs can be employed to enhance the intelligence of drill-down processes within visual analytics systems. Researchers face mainly three key challenges. Firstly, during the drill-down process, the decision space expands dramatically, creating a combinatorial explosion that makes it increasingly difficult for users to identify and prioritize valuable drill-down paths. Secondly, drill-down operations often fail to effectively capture and align with user intent, resulting in exploration directions that may be irrelevant or only partially useful. Lastly, the lack of systematic management of the drill-down context leads to fragmented and disorganized exploration histories, which in turn diminishes the readability and continuity of the overall analytical process.


To address these challenges, this paper introduces Intelligent Drill-Down, a large language model-driven drill-down technique for human-AI collaborative visual exploration. By combining the capability of the LLM with user requirements, the system analyzes and identifies drill-down paths worth exploring, and further assists users in selecting appropriate drill-down directions through analytical visualizations. Specifically, we propose a drill-down path recommendation method, in which the LLM is trained to approximate a validated greedy algorithm. Secondly, we integrate users’ natural language inputs with their interaction histories to infer user intent, which is then used to construct drill-down charts. Finally, we generate multiple insights for drill-down and visualization by combining user requirements with visual analytics. Building upon this framework, we design a system that features a hybrid interface, which incorporates hierarchical navigation to track users’ exploration and manage parallel branches, a visualization panel for interactive data exploration, and an insight panel to present analytical findings and provide drill-down recommendations. Our main contributions are as follows:
\begin{itemize}
    \item{We propose an intelligent drill-down framework, using LLM based visualization insights and drill-down path recommendations. This approach reduces the burden on users and helps them drill-down along valuable paths.} 
    \item{We develop a system that embodies our framework, supporting users to drill-down with the assistance of agents and providing them with multiple interaction methods and parallel exploration support.}
    \item{We demonstrate the effectiveness of the system through a usage scenario and a user study.} 
\end{itemize}

\section{Related Work}

In this section, we review existing literature regarding user intent, visualization generation and drill-down.

\subsection{Recommendation based on User Intent}
Dimara et al.~\cite{dimara2019interaction} summarize user intent in the visualization domain as encompassing a goal, task, or problem that users seek to address. In practice, these intents may manifest in diverse forms, such as exploring datasets and charts~\cite{lee2012beyond}, extracting meaningful insights~\cite{gotz2009characterizing}, or guiding subsequent analytical actions~\cite{abras2004user}. To support these intents, researchers have increasingly turned to machine learning and, more recently, large language models (LLMs) to better understand, infer, and even anticipate user needs.

For instance, systems like AutoCaption~\cite{liu2020autocaption} and ChartInsighter~\cite{wang2025chartinsighter} automatically generate natural language descriptions for visualizations, lowering the cognitive burden on users and assisting them in deriving insights more effectively. Li et al.~\cite{li2022diverse} employ an LSTM-based model to recommend interaction strategies that optimize user exploration within visualization systems. LEVA~\cite{zhao2024leva} leverages LLMs to recommend tasks that are most relevant to ongoing exploration patterns, thereby personalizing the analytical workflow.
Recent advances have further shifted from reactive support to proactive LLM-driven agents. For example, ProactiveVA~\cite{zhao2025proactiveva} provides users with context-aware assistance by monitoring user interactions, aiming to give users suggestions when most needed.

However, the studies mentioned above primarily aim to enhance the general VA workflow, with limited emphasis on the concept of drill-down. Our work conducted an in-depth investigation into the context of drill-down human-computer collaboration.




\subsection{LLM for Visualization Generation}
LLM-based visualization generation has emerged as a frontier research direction at the intersection of natural language processing and visual analytics. Current research efforts primarily focus on transferring natural language into corresponding visualization specifications, enabling users to express analytical needs more intuitively.

In the early stages, studies focus on the automatic generation of individual charts. For example, CHAT2VIS~\cite{maddigan2023chat2vis} pioneers the use of LLMs to produce visualization code directly from textual input and conducts systematic comparisons across different LLMs to evaluate their performance. Building on this foundation, subsequent systems such as LIDA~\cite{dibia2023lida} and ChartGPT~\cite{tian2024chartgpt} extend the paradigm by leveraging prompt engineering and fine-tuning strategies, respectively. Further refinements are introduced by Li et al.~\cite{li2024visualization}, who demonstrate that incorporating one-shot and few-shot learning approaches substantially enhances the accuracy and robustness of visualization generation, particularly in complex or ambiguous cases.

As the field evolves, researchers begin to focus on generating linked views through task decomposition. For instance, SmartMLVs~\cite{qiu2025smartmlvs} employs Retrieval-Augmented Generation (RAG) techniques to generate multiple interactive views through a structured three-step process of task decomposition, visualization generation, and cross-view linking. Similarly, LightVA~\cite{zhao2024lightva} emphasizes dialogue-driven task decomposition, supporting users in iterative and multi-round explorations where visual outputs adapt dynamically to evolving analytical questions. Previous studies, however, do not consider drill-down operations within the context of visual analysis.

Our work primarily focuses on leveraging the LLM to generate recommended drill dimensions and visual insights, thereby alleviating users’ cognitive load during path exploration in the drill-down process.

\subsection{Drill-Down Techniques}
Research on intelligent drill-down aims to overcome the limitations of traditional \textcolor{black}{Online Analytical Processing (OLAP)~\cite{chaudhuri1997overview}}, which often overwhelms analysts with redundant details. Smart Drill-Down~\cite{joglekar2017interactive} introduces rule ranking and dynamic sampling to highlight the most interesting subsets and maintain interactive performance on large tables. In the education domain, Automated Insight Drill-Down algorithms~\cite{khosravi2021intelligent} recommend meaningful student subgroups for closer inspection, guiding teachers to focus on anomalous patterns. Similarly, provenance-based approaches~\cite{ikeda2012provenance}, such as the Panda system, enable fine-grained drill-down into workflows by tracing how outputs derive from inputs, thereby providing deeper transparency in complex data pipelines.

At the same time, other studies focus on improving the validity and scalability of drill-down paths. VisPilot~\cite{lee2012beyond} identifies the drill-down fallacy—misinterpreting local differences as causal—and proposes guided paths that ensure safety, saliency, and succinctness. To reduce the exponential growth of possible drill-down paths, pruning strategies eliminate uninformative operations~\cite{zhang2007effective}. Beyond general data analysis, extensions include multi-foci drill-down~\cite{conklin2002multiple} across tuple and attribute hierarchies, and deep drill-down~\cite{nemeth2021deep} for industrial fault detection. However, the existing research remains at a relatively early stage.

Our work fully leverages the capabilities of LLMs for path exploration, code generation, and insight derivation, with the goal of developing a more intelligent drill-down framework.
\section{Intelligent Drill-Down Framework}
The drill-down pipeline consists of three stages: selection, generation, and analysis~\cite{vartak2014seedb,wongsuphasawat2017voyager}. Therefore, in Intelligent Drill-Down, our goal is to reduce the interpretation burden on analysts and to intelligently recommend drill-down paths, helping analysts more easily drill-down according to their intentions. In the following section, we describe the challenges and derive the design requirements for integrating the LLM into the user’s workflow. Finally, we introduce the conceptual framework of an LLM-driven drill-down.

\begin{figure}[htbp]
    \centering
    \includegraphics[width=0.85\linewidth]{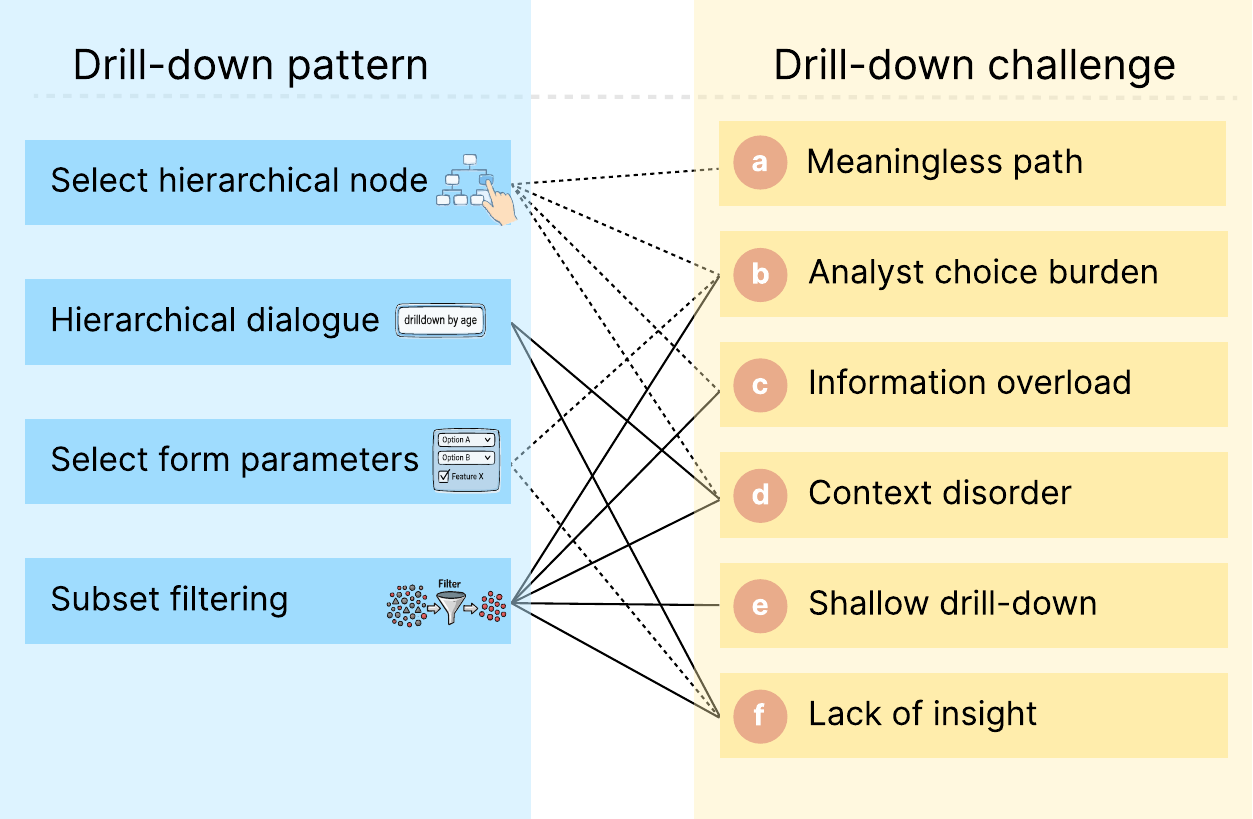}
    \caption{\textcolor{black}{Mapping between drill-down patterns (interaction modes) and associated challenges. The lines identify which interaction behaviors lead to specific challenges (line styles alternate solely for visual clarity).}}
    \label{drill-down pattern & problem}
\end{figure}




\subsection{Challenges}
\textcolor{black}{Through a literature review, we categorize common drill-down patterns (i.e., existing interaction modes) and analyze the specific challenges they induce, as mapped in Figure  \ref{drill-down pattern & problem}.} The following challenges are common issues of drill-down and are independent of the domain.





\textbf{C1: Expansive drill-down exploration space and heavy analyst choice burden.}
Complex and high-dimensional data structures allow analysts to explore information from multiple perspectives~\cite{stolte2002polaris}.
To extract valuable insights without overlooking critical findings, analysts must therefore navigate an overwhelming array of options and potential outcomes~\cite{sarawagi1998discovery}, including numerous combinations of filters, groupings, encodings, and drill-down paths that rapidly expand the choice space \textcolor{black}{(Fig. \ref{drill-down pattern & problem}b)} 
~\cite{zhang2007effective,wongsuphasawat2015voyager}.
Making accurate judgments in this complexity not only heavily depends on analysts’ expertise in data interpretation but also imposes a substantial burden in terms of interpretation and selection, ultimately resulting in information overload \textcolor{black}{(Fig. \ref{drill-down pattern & problem}c) }~\cite{sarawagi2000user,khosravi2021intelligent}.
\textcolor{black}{Consequently, such analysis paralysis risks overlooking critical data insights, leading to suboptimal or erroneous decisions.}


\textbf{C2: Meaningless path and user intent misalignment.} 
Repeated drilling into the same analytical dimension may yield diminishing returns, resulting in a shallow drill-down that fails to uncover deeper insights  \textcolor{black}{(Fig. \ref{drill-down pattern & problem}e)} ~\cite{joglekar2017interactive}. 
The generation of similar or redundant visualizations offers limited analytical value ~\cite{mafrur2018dive}, leading users to expend excessive effort and time on meaningless path \textcolor{black}{(Fig. \ref{drill-down pattern & problem}a)}. 
\textcolor{black}{Furthermore, rigid interactions (e.g., clicking) lack the expressiveness to convey complex analytical goals, forcing users into trial-and-error loops to locate desired insights~\cite{siddiqui2016effortless,shanmugasundaram1999compressed}.}
\textcolor{black}{Such inefficiency prolongs time-to-insight, significantly hampering overall analyst productivity.}

\textbf{C3: Fragmented drill-down context and impaired readability.}
As users drill-down to deeper hierarchical levels, they are often required to frequently navigate between different levels of visual representations to conduct relational analysis~\cite{stolte2002polaris}.
This imposes a high cognitive load, reduces interpretability, and often leads to user disorientation during the drill-down process \textcolor{black}{(Fig. \ref{drill-down pattern & problem}d)} ~\cite{lam2008framework}.
\textcolor{black}{This fragmentation isolates details from the global context, leading to misinterpretations and erroneous conclusions~\cite{lee2019avoiding}.}

\begin{figure*}
    [htbp]
    \centering
    \includegraphics[width=0.83\linewidth]{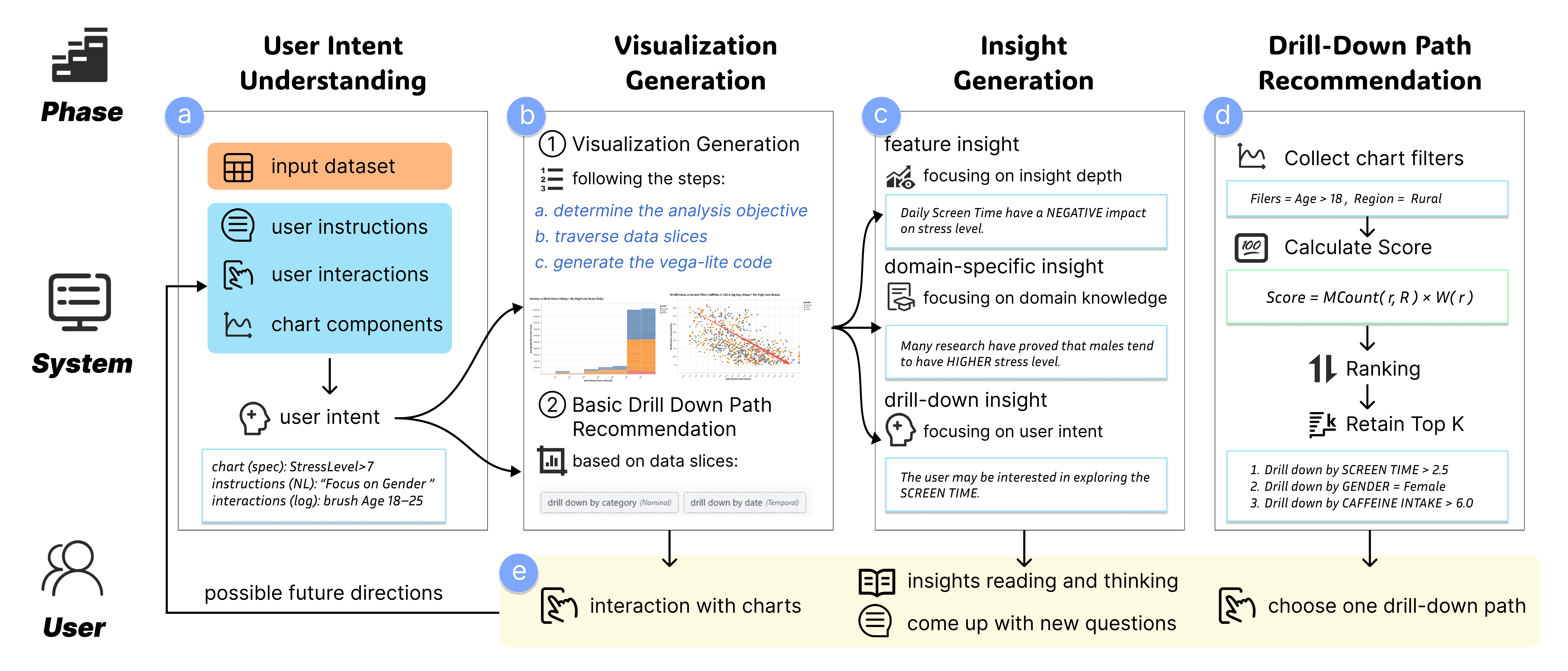}
    \caption{The workflow of the Intelligent Drill-Down. It mainly includes four stages: user intent understanding (a), visualization generation (b), \textcolor{black}{insight generation (c) and drill-down path recommendation (d)}. Users can start the new iteration with obtained insights and recommendations in the latter three stages (e).}
    \label{framework}
\end{figure*}

\subsection{Design Requirements}
Based on these challenges, we summarize the following design requirements.

\textbf{DR1: Intelligent generation of drill-down visualizations:}
The system should support the intelligent generation of drill-down visualizations that facilitate user data analysis. This involves accurate data identification and transformation, as well as providing a diverse range of interactive options to promote thorough exploration. Furthermore, visualizations should be capable of interpreting user intent, selecting appropriate filters, and presenting fine-grained views aligned with analytical objectives. \textbf{(C1, C3)}

\textbf{DR2: Support for branch management:}
The system ought to embed intelligent branch management to facilitate exploration from multiple viewpoints. This includes support for parallel exploration through multiple drill-down dimensions and efficient switching between branches. In addition, a streamlined management interface should be provided to minimize confusion and maintain clarity during the exploration process.\textbf{(C3)}

\textbf{DR3: Automated generation of insights and drill-down options:}
To reduce the cognitive load associated with interpreting visualizations, the system should automatically generate succinct analytical summaries based on the user's drill-down path, thus supporting the interpretation of visual content \textcolor{black}{(Fig. \ref{drill-down pattern & problem}f)}. 
To ease the decision-making burden in selecting exploration paths, the system should also provide a set of contextually relevant drill-down options that users can either execute directly or consult as guidance for further exploration. \textbf{(C1, C2)} 

\textbf{DR4: Clear representation of drill-down paths:}
The system should explicitly surface the available drill-down paths and render them in a structured, human-readable form, including historical steps and applied filters, to help users understand their current position within the exploration process, verify which data subsets are in effect, and quickly roll back or revisit earlier decisions when necessary. This transparency prevents disorientation during analysis and supports informed decision-making for subsequent drill-down actions. \textbf{(C1, C3)}

\textbf{DR5: Natural language-driven drill-down and interaction capture:}
\textcolor{black}{The system should log key user interactions and incorporate natural-language requests as a complementary intent signal. Combining these explicit and implicit cues helps disambiguate high-level goals that clicks alone may not reveal and steers drill-down recommendations away from low-yield paths.\textbf{(C2)}}

\subsection{Conceptual Framework}
Based on the identified challenges and design requirements, we propose Intelligent Drill-Down.
Given the capabilities of the LLM in comprehending abstract analytical intents~\cite{zhao2024leva,gao2023text}, invoking external tools~\cite{yao2022react,gao2023pal}, and leveraging domain-specific knowledge~\cite{luo2022biogpt}, our framework utilizes the LLM to translate high-level analytical goals into concrete drill-down operations and corresponding visualization code, which is superior to traditional machine learning models~\cite{wu2024automated,maddigan2023chat2vis}. 


Our framework consists of four stages: intention understanding, visualization generation, insight generation, and drill-down path recommendation, as shown in Figure~\ref{framework}. Users interact with the visualization and then initiate a drill-down either by selecting from the “recommended drill dimensions” tag or by submitting a query in natural language. The LLM subsequently interprets user intent to generate a visualization that best reflects the drill-down context, together with three basic candidate dimensions for further exploration. Finally, users receive analytical and drill-down insights derived from the current visualization, as well as three high-level recommended directions for subsequent exploration. In the latter three stages, users can directly choose a drill-down direction based on the obtained insights and recommendations combined with their own judgment, thereby initiating a new iteration \textcolor{black}{(Fig. \ref{framework}e)}.

\textbf{Intention Understanding:}
To generate a visualization that accurately reflects the user's analytical intent, we propose decomposing the intent along three dimensions \textcolor{black}{(Fig. \ref{framework}a)}. 
Specifically, we instruct the LLM to analyze the user's analytical needs based on: chart component, employed to infer users' analytical needs from their previous drill-down activities, 
the user's interaction history and the user's drill-down instructions, including either natural language input or the graphical interface interaction.

\textbf{Visualization Generation:}
We aim to continue the drill-down from the current analytical level based on the user intent inferred in the previous stages, and generate corresponding visualization charts \textcolor{black}{(Fig. \ref{framework}b)}.
We employ Vega-Lite as the visualization specification language. Vega-Lite is a lightweight framework designed for the generation of interactive visualizations~\cite{satyanarayan2016vega}. When selecting chart types, we instruct the LLM to choose the type that effectively reveals the salient features of the data, while ensuring that the visualization supports rich interactive functionalities such as brushing and hovering. Additionally, we request the LLM to select several basic recommended drill-down dimensions available for the user to click for drill-down or reference by evaluating both intent alignment and redundancy minimization.




\textbf{Insight Generation:}
Upon the user selecting “generate the insight,” we aim to produce three categories of insights related to the current layer's visualization: data feature insights, domain-specific insights, and drill-down insights \textcolor{black}{(Fig. \ref{framework}c)}.
The LLM is instructed to assign differential weights to each insight type based on its analytical depth, prioritizing those that are more effective in addressing the user's underlying information needs. 

\textbf{Drill-Down Path Recommendation:}
Concurrently, when the user selects “generate the insight”, we generate three high-level recommended drill-down dimensions to support further exploratory analysis \textcolor{black}{(Fig. \ref{framework}d)}.
We consider drill-down paths that yield greater information gain and are better aligned with user intent to be high-value drill-down paths. Concretely, we introduce a modified greedy algorithm inspired by smart drill-down~\cite{joglekar2017interactive}, in which the LLM computes a drill-down relevance score for each potential drill dimension. The top three dimensions with the highest scores are selected as the recommended drill-down directions, replacing the basic candidates generated in the earlier stage. 
In our work, we propose two types of drill-down dimensions: basic drill dimensions and high-level drill dimensions. The basic drill dimensions are generated in tandem with the visualization process. These dimensions are derived from data slicing and, by incorporating the current Vega-lite code and user intent, identify suitable drill-down paths. Although this type of recommendation is relatively coarse, it is generated concurrently with the visualization process, requiring no additional computation time. It can still offer users helpful guidance and actionable suggestions. In contrast, high-level drill dimensions are generated concurrently with the derivation of insights. This process entails additional computational overhead but yields more accurate and targeted drill-down recommendations. The system supports both types of drill dimensions, enabling users to flexibly select the appropriate mode based on their analytical requirements.


\section{Intelligent Drill-Down System}
Based on the framework above, we propose the Intelligent Drill-Down system, \textcolor{black}{which integrates LLM-based prompt engineering with rule-based logic}.
Our system consists of four components: (1) intention understanding and visualization generation; (2) insight generation and drill-down path recommendation; (3) hierarchy navigation; and (4) a user interface.


\label{sec :intention understanding and visualization generation}
\subsection{Intention Understanding and Visualization Generation}
This section presents the specific methodologies for intention understanding and visualization generation, and elaborates on how they mutually influence each other.


\subsubsection{Intention Understanding}
To generate visualizations aligned with users' drill-down intentions, the system infers analytical intent by integrating three types of signals \textbf{(C2)} (Fig. \ref{framework}a). First, it \textcolor{black}{extracts filtering conditions from the transform-filter fields in the current Vega-Lite specification}
the user has set as an explicit drill-down path. These base filters directly reflect the data subset the user is focusing on~\cite{chaudhuri1997overview}.

Second, to infer analysis goals, our system logs interactions in a semantic schema: (action\_type, target\_fields, predicate, value\_range). 
\textcolor{red}{Drawing on Vega-Lite’s grammar ~\cite{satyanarayan2016vega}, the predicate is defined as a set of logical conditions mapping data tuples to Boolean values (True/False) to determine their membership in the current selection set. Instead of recording raw pixel coordinates, the system translates physical interactions into explicit semantic filters; for instance, a scatterplot brush is captured as a precise multidimensional range predicate (e.g., (Income\( >= \)100k) AND (Age\( <= \)30)). This abstraction ensures chart independence, as the intent is preserved regardless of the visual encoding. Detailed examples of these interaction-tracking patterns across various chart types are provided in Table 1 of the Supplemental Material.}
\textcolor{black}{Instead of using a separate classifier, we inject these structured logs into the LLM prompt alongside distilled action–intent mappings (e.g., filter $\to$ constrain ~\cite{gotz2009characterizing,brehmer2013multi}) and instructions to prioritize high-frequency and recent actions. In a single invocation, the model processes these inputs to output inferred task hypotheses followed by the next Vega-Lite specification.}

Third, our system records and parses drill-down instructions including either natural language input or the selection of recommended drill dimension tags, extracting any stated analysis interests, relevant data dimensions, and target objectives. For instance, if the user types a query about a certain metric or requests a drill-down via text, \textcolor{black}{the system} interprets it as a corresponding filter or analytical operation. 

These three signals—explicit visualization settings, implicit interaction cues, and explicit language descriptions—are jointly leveraged.
\textcolor{black}{Crucially, this integration ensures robustness; in cold-start scenarios lacking interaction history, the system prioritizes explicit settings and descriptions to derive intent.}
By combining \textcolor{black}{available} behavioral cues with explicit user inputs, our approach infers user intent more comprehensively.

\subsubsection{Visualization Generation}
After the user selects the drill-down operation, the system transforms “intent” into “drill-down visualization” through the following pipeline (Fig. \ref{framework}b).


\textbf{Drill-Down Construction:} 
\textcolor{black}{The system initializes the drill-down by parsing the current Vega-Lite specification to preserve the user's established context. To construct the new view, the LLM maps the three inferred intent signals into filter predicates: (1) explicit transform-filter fields are retained as base constraints; (2) interaction logs (e.g., specific selection bounds) are translated into range or categorical filters; and (3) natural language instructions are parsed to target specific dimensions or values. These new predicates are appended incrementally to avoid conflicts. Finally, the updated specification undergoes a two-stage validation: structural validation against the Vega-Lite schema and semantic validation to ensure field existence and data type consistency. In the event of a validation or rendering failure, the system triggers a rollback mechanism.}
\textcolor{red}{Specifically, when the system captures a frontend rendering exception during a drill-down operation, it automatically reverts the UI to the pre-operation state to ensure stability. Simultaneously, the caught error stack trace is extracted and fed back to the LLM as a corrective prompt to regenerate the executable code.}

\textbf{Task-Semantic Visualization Selection and Interaction Availability:} In order to enhance the drill-down expressiveness, the system chooses the most appropriate visual form to match the user’s task semantics and supports multiple modes of interaction \textbf{(C2)}. 
\textcolor{black}{To ensure design validity and mitigate generation instability, we codify established visualization heuristics~\cite{cleveland1984graphical,mackinlay1986automating} (e.g., mapping temporal trends to line charts) into the prompt as strict constraints. This directs the LLM to deduce the chart type based on the identified task and data types, rather than relying on opaque parametric knowledge. Furthermore, to guarantee functional interactivity, we provide few-shot demonstrations of Vega-Lite’s declarative syntax (e.g., defining an interval selection for brushing). This context guides the LLM to accurately configure selection parameters that enable coordinated cross-view filtering and context preservation, effectively closing the loop of intent, interaction, and recommendation.}


\textbf{Basic Recommended Drill Dimensions:} 
\textcolor{black}{Given the updated spec and interaction context, the system generates a candidate pool and leverages the LLM's semantic reasoning capabilities to rank dimensions based on two strict prompt constraints:}

\textcolor{black}{(1) \textbf{Intent Alignment:} Evaluating the semantic similarity between dimensions and the user’s drill-down instructions.}

\textcolor{black}{(2) \textbf{Redundancy Minimization:} Checking for logical inclusion with current filters to ensure orthogonal perspectives and avoid overlaps.}

Although these recommended drill dimensions are somewhat preliminary, they still offer users potential directions for further drill-down and help alleviate their cognitive workload \textbf{(C1)}. 
Redundancy minimization is designed to offer users diverse drill-down paths, thereby providing richer exploration references and mitigating the issue of shallow drill-down \textbf{(C2)}.


\begin{figure*}
    [htbp]
    \centering
    \includegraphics[width=0.83\linewidth]{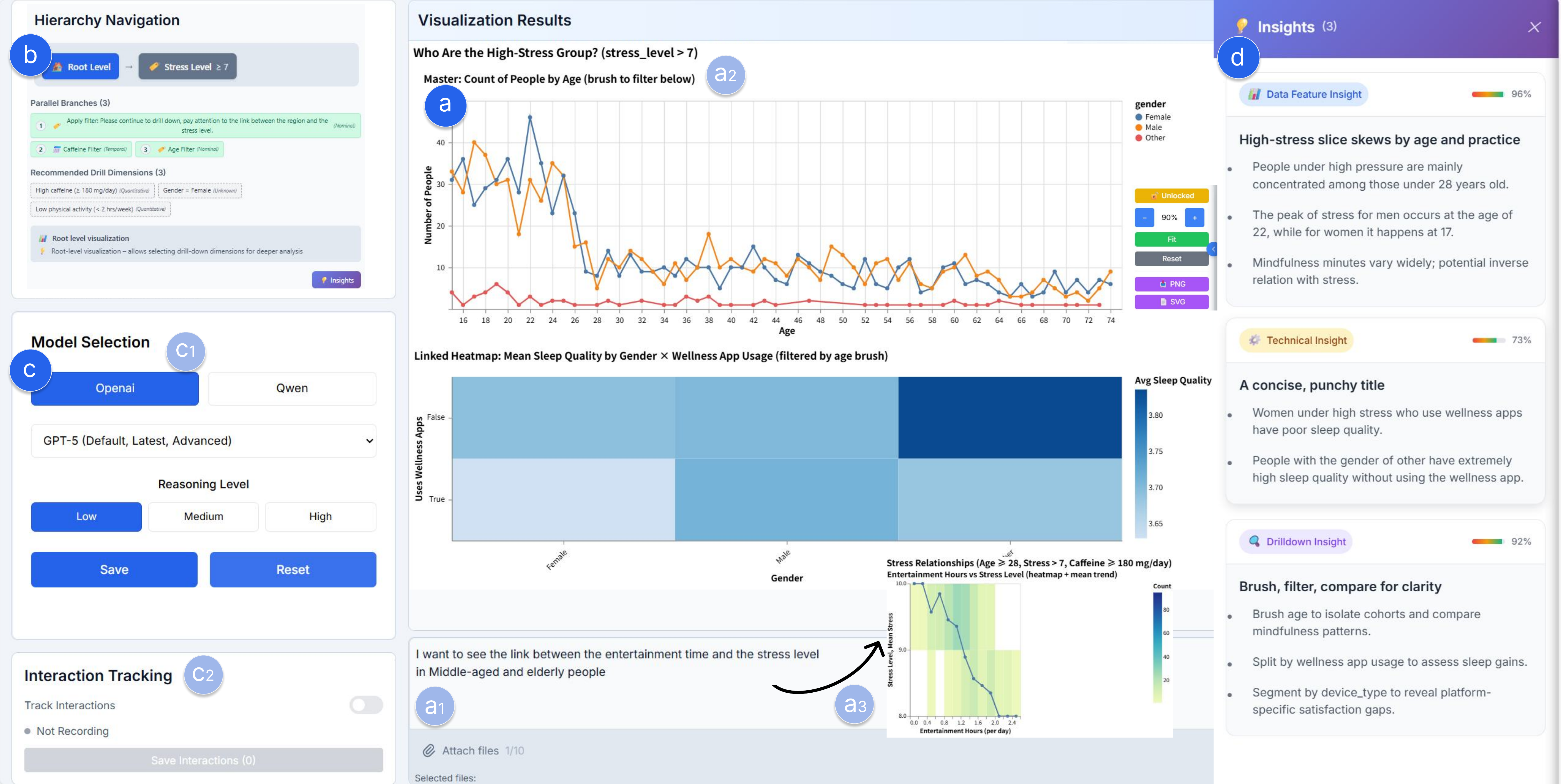}
    \caption{
    \textcolor{black}{The Intelligent Drill-Down system mainly comprises 4 components. (a) Users upload datasets and instructions via (a1) to receive visualizations in (a2); (a3) illustrates the drill-down process. (b) Path navigation (breadcrumb trail) and parallel controller for branch management. (c) Configuration for model selection (c1) and interaction tracking (c2). (d) The insight panel for data-driven decision support.}
    }
    \label{system}
\end{figure*}


\subsection{Insight Generation and Drill-Down Path Recommendation}
Once the user selects "generate insight", the system will concurrently generate both the insights \textcolor{black}{(visualized in Fig. \ref{system}d)} and the drill-down path recommendation \textcolor{black}{(Fig. \ref{system}b)}.

\subsubsection{Insight Generation}
To reduce the cognitive load on users in interpreting visual data and selecting appropriate drill-down paths for further analysis, the system automatically generates insights categorized into three groups: Data Feature Insights, Domain-Specific Insights, and Drill-down Insights \textbf{(C1)} (Fig. \ref{framework}c) ~\cite{demiralp2017foresight}. Data Feature Insights focus on patterns and characteristics inherent in the data itself – for example, discoveries driven by the visualization such as extreme values, value ranges, emerging trends, or comparisons between groups evident in the current view. 
Domain-Specific Insights \textcolor{black}{ (dynamically instantiated as 'Technical', 'Business', or 'Clinical' insights depending on the data context)} incorporate domain knowledge to interpret the data patterns, bringing in external knowledge or expertise  to help the user understand the visualization results from a professional perspective ~\cite{li2025role}. 
\textcolor{black}{Leveraging the LLM's inherent domain expertise ~\cite{singhal2025toward,achiam2023gpt}, we prompt the model to interpret detected visual patterns (e.g., trends) using its internal knowledge base, directly translating data features into professional explanations.}
Drill-down Insights highlight cues for further exploration, often pointing out interesting phenomena that emerge when examining a subset of the data or suggesting the next steps for analysis. To generate these insights, each category follows a four-step pipeline:



\textbf{(a) Visualization Analysis:}
The system prompts the LLM to denoise and quantize the colors of the current visualization. Subsequently, the LLM is instructed to adopt a classification and data extraction approach to identify noteworthy data-driven patterns~\cite{savva2011revision}, including outliers, distribution ranges, trend directions, and inter-group contrasts.
\textcolor{black}{Guided by prior studies~\cite{wang2025chartinsighter}, we prompt the LLM to score each insight using an explicit rubric: magnitude of deviation and consistency. We further instruct the model to prioritize multidimensional trend relationships, assigning them higher scores due to their utility in revealing complex, non-obvious patterns.}
This step produces an initial score for each potential insight based on the significance of the detected pattern in the data.

\textbf{(b) User Intent Alignment:}
\textcolor{black}{We compute an alignment flag for each insight by checking if its involved data fields or value ranges overlap with the user’s inferred intent signals (e.g., predicates from interactions or entities from natural language queries). Aligned insights are promoted to a higher-priority tier; ties within each tier are broken by the initial significance score. This re-ranking explicitly favors insights consistent with the user’s current focus.}

\textbf{(c) Scoring and Filtering:} 
\textcolor{black}{The system (not the LLM) performs the final ranking deterministically. 
For each candidate insight $i$, we extract a visualization-based significance score $S_{vis}(i)$ from the LLM's structured output (step a) and compute a binary intent-alignment indicator $I_{align}(i) \in \{0, 1\}$ via rule-based field/range overlap checking (step b). 
We then rank the insights using the lexicographic key $(I_{align}(i), S_{vis}(i))$ in descending order (i.e., aligned insights always precede unaligned ones, and ties are broken by $S_{vis}$). 
Mathematically, this is equivalent to computing a total score $S_{final}(i) = S_{vis}(i) + \lambda \cdot I_{align}(i)$, where $\lambda$ is a constant set to $S_{vis}^{max} + 1$. 
Finally, the system filters the results to retain only the top-ranked insights for presentation.}

\subsubsection{Drill-Down Path Recommendation}
\label{sec:drill-down path}
To assist users in identifying the next step for deeper analysis, our system employs and modifies a greedy scoring strategy inspired by the “Smart Drill-Down” method to evaluate the information gain of candidate drill-down rules \textbf{(C1)} (Fig. \ref{framework}d) ~\cite{joglekar2017interactive}. Compared to the CoT and Beam search algorithms, the greedy algorithm demonstrates higher computational efficiency~\cite{won2023break,lee2025well}, thereby reducing user waiting time and improving the coherence of the drill-down process.
Specifically, the incremental score for a candidate rule is defined as:
\begin{equation}
    \label{Score}
    Score (r|R) = MCount(r,R) \times W(r)
\end{equation}
In this equation, \(MCount(r,R)\) represents the number of the new data groups covered by a new rule \(r\) that are not already covered by the current set of rules \(R\), in other words, the additional information contributed by \(r\) given \(R\). \(W(r) = \sum_{f\in r} \alpha_f\times\log_2(|f|)\) is a weight reflecting the complexity of rule  \(r\) . \textcolor{red}{Here, \(|f|\) denotes the domain size of attribute \(f\), that is, the number of  distinct values for field \(f\),} \(\log_2(|f|)\) quantifies the complexity contributed by filtering on \(f\) and \(\alpha_f\) is a coefficient that modulates this contribution, \textcolor{black}{quantified based on the semantic relevance between attribute \(f\) and the user's analysis intent (parsed by the LLM).}


In implementation, \textcolor{black}{the system leverages} the LLM to extract the base filters of the current view as the initial rule set \(R\). \textcolor{black}{Subsequently, the system} enumerates all feasible candidate drill-down rules by considering dimensions not yet used in the current drill-down path and their possible filter values.
Giving priority to unused dimensions can expand data coverage and promote diversity~\cite{wongsuphasawat2016towards,wongsuphasawat2017voyager}.
For example, if the current view is already filtered by region and year, then other dimensions (such as product category or customer segment) and their values would form the pool of candidate drill-down rules. 
\textcolor{black}{Next, for each candidate rule $r$, the system executes queries to determine its marginal coverage $\mathit{MCount}(r, R)$ relative to the current selection $R$, and calculates its complexity weight $W(r)$ using the derived $\alpha_f$, thereby obtaining $\mathit{Score}(r|R)$.}
The candidates are then ranked by this score, and a greedy strategy is applied to select the top-k drill-down paths. 
After each path is selected, the \(Score (r|R)\) values for the remaining candidates are recomputed with respect to the updated rule set \(R \cup R'\) before the next selection, and the process repeats until \(k\) paths have been chosen. Through this stepwise optimization, the recommended drill-down paths are expected to offer the greatest information gain from the current data slice.


From a time complexity perspective, let \(n\) be the number of candidate drill-down rules and \(k\) be the number of paths to recommend. The above scoring and greedy selection process runs in approximately \(O(n \times k)\) time. \textcolor{red}{The breakdown of per-iteration end-to-end latency, LLM calls, and candidate counts is shown in Table 2 of the supplementary material.}
Further discussion is provided in Sec. \ref{sec: Performance of LLMs}.

\subsection{Hierarchy Navigation}

\textcolor{red}{While the underlying exploration process constructs a complex directed acyclic graph (DAG) of analysis states, visualizing this full structure as a traditional node-link tree often introduces significant visual clutter and cognitive overhead, particularly within the constrained space of an analytics sidebar ~\cite{heer2008graphical}. To address the \textbf{challenge of context management (C3)} without overwhelming the user, we decouple the visualization of the exploration hierarchy into two distinct, coordinated interface components: \textbf{Path Navigation} for the active linear lineage, and a \textbf{Parallel Controller} for managing branching hypotheses. This hybrid design pattern prioritizes the readability of the current analytical context while maintaining accessible pathways to alternative lines of inquiry.}
\subsubsection{Path Navigation}
\textcolor{black}{To visualize the hierarchical structure without visual clutter, the interface renders the current active branch of the exploration tree as a linear breadcrumb trail (see Fig. \ref{system}b, top)}.
This trail explicitly enumerates the ordered sequence of drill-down operations from the root view to the user’s current visualization. 
\textcolor{black}{Each filter step is displayed as an interactive token.}
\textcolor{black}{Selecting any breadcrumb initiates a jump-to-state operation by retrieving the node’s full Vega-Lite specification. This specification includes the data state—defined as the cumulative filter transforms and data source configuration—alongside the visual encoding.}
\textcolor{black}{This design supports rapid vertical backtracking. To navigate the tree horizontally or extend it, users interact with the Parallel Branches (to switch contexts) and Recommended Drill Dimensions (to add child nodes) shown in Fig. \ref{system}b, effectively managing the full drill-down hierarchy.}

\begin{figure*}
    [htbp]
    \centering
    \includegraphics[width=0.83\linewidth]{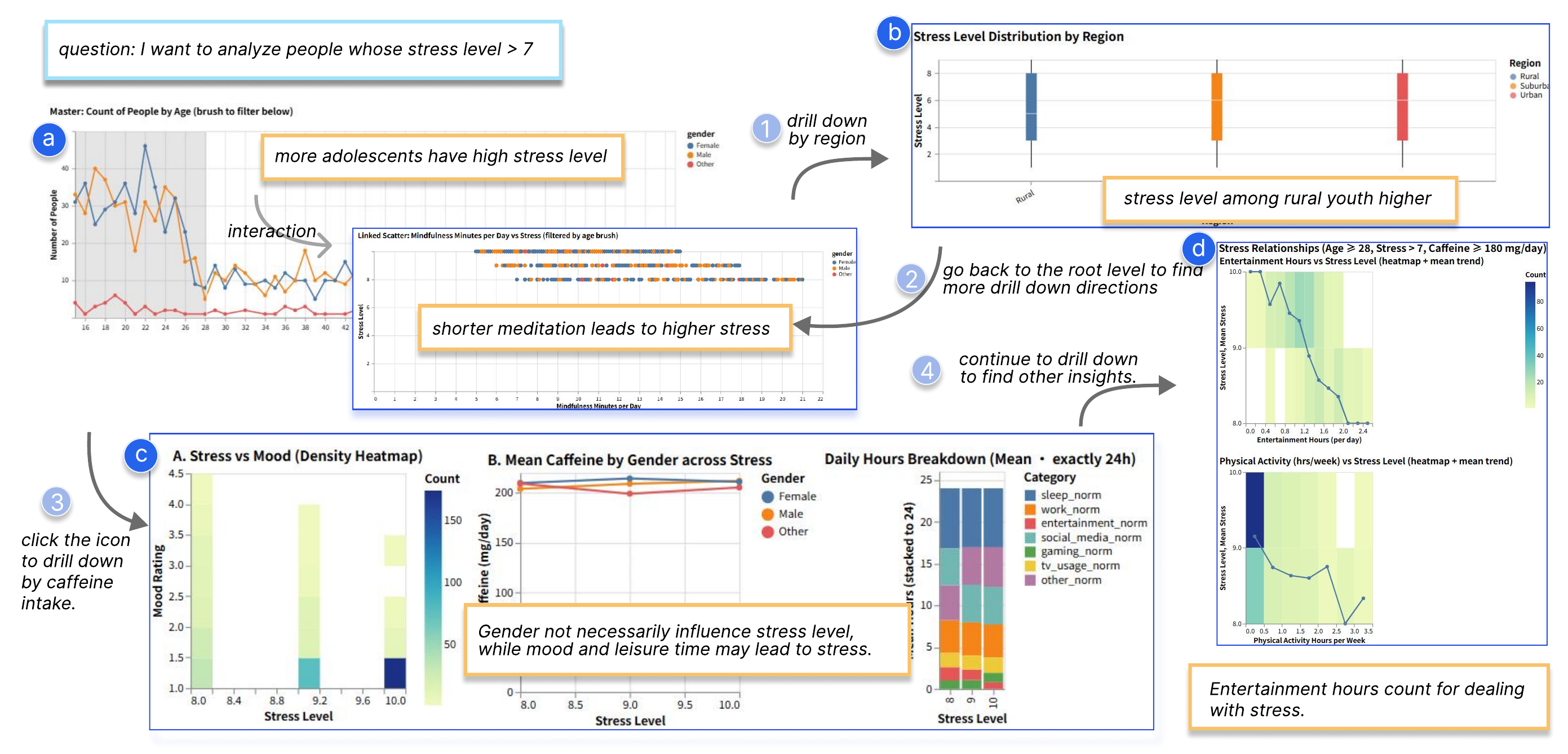}
    \caption{A usage scenario to demonstrate how Intelligent Drill-Down operates. \textcolor{black}{Blue boxes denote user inputs, and orange boxes highlight insights discovered during the exploration. The analysis follows the numerical path: (1) Drilling down from the initial overview (a) to a regional view (b) via natural language; (2) backtracking to the root level to switch exploration branches; (3) selecting a recommended dimension (caffeine) to reveal multivariate factors in (c); and (4) further drilling down to detailed correlations in (d).}}
    \label{usage scenario}
\end{figure*}

\subsubsection{Parallel Controller}

\textcolor{red}{To support non-linear exploration where analysts may wish to pursue multiple competing hypotheses simultaneously, the system uses a  Parallel Controller (Fig. \ref{system}b). Instead of forcing users to commit to a single analytical path or manually managing complex history trees, this component simplifies branch management into a list-based selection interface of complete root-to-leaf paths, each distinctively labeled by its final leaf node. When a user initiates a new drill-down from a visited node, the system implicitly forks the exploration tree, generating a new branch independent of prior paths.}

\textcolor{red}{The Parallel Controller visualizes these branches as distinct exploration contexts, enabling users to "park" one line of inquiry and switch to any branch from the global pool with a single click. This mechanism eliminates the cognitive friction of manual rollback-and-replay; users can fluidly compare disparate subsets (e.g., comparing  Rural vs. Urban trends in parallel) by toggling between branches. By abstracting the complexity of the underlying branch history into a streamlined list view, the system encourages broad exploration and hypothesis pruning while maintaining a focused workspace ~\cite{terry2002side}.}


\subsection{Interface}
We design an interface to bridge users and our system. The interface comprises four main components: I/O Interface, Hierarchy Navigation, Main Toolbar and Insight Panel, as shown in Figure \ref{system}.

\textbf{I/O Interface: }
The I/O interface serves as the gateway for data input and visualization output (Fig. \ref{system}a). Users can upload up to ten data files through this component and specify their drill-down intentions using natural language (Fig. \ref{system}-a1). Based on these inputs, the system produces an interactive visualization that supports user interactions such as scroll-based zooming for inspecting details (Fig. \ref{system}-a2). A small toolbar on the right side of the visualization provides additional controls, such as exporting the current chart as a PNG image and locking the current interface state.

\textbf{Hierarchy Navigation:} The hierarchy navigation module helps users track and explore the multi-level drill-down process and comprises three parts: path navigation, a parallel branch controller, and recommended drill dimensions (Fig. \ref{system}b). The path navigation bar displays the current drill-down path as a breadcrumb trail, enabling users to roll back by clicking on any breadcrumb to return to a previous level. The parallel branch controller lists all parallel exploration branches and allows users to switch between different branches of analysis. The recommended drill dimensions section presents up to three system-suggested drill-down options, which users can click to directly navigate into those deeper, system-suggested analyses.

\textbf{Main Toolbar:} The main toolbar provides global controls for the system, including model selection (Fig. \ref{system}-c1) and interaction tracking (Fig. \ref{system}-c2). Here, users can choose the backend model for analysis (e.g., GPT-4o, GPT-5, GPT-5 mini or Qwen-Long) and set the reasoning level (low, medium, or high) to control the complexity and depth of the system’s analysis. Via the main toolbar, our system offers a variety of LLMs, enabling users to select from different models. The performance differences among these LLMs are discussed in \textcolor{red}{Sec. \ref{sec: Performance of LLMs}}.
The toolbar also includes a reset function and a toggle for interaction tracking. When interaction tracking is turned on, the system leverages the user’s past interactions to generate visualizations that more closely align with the user’s overall analytical objectives, thus helping users reach their intended insights more efficiently.

\textbf{Insight Panel:} The insight panel presents automatically generated insights from multiple perspectives to help users interpret the current drill-down visualization (Fig. \ref{system}d). This panel displays three types of insights: data feature insights, domain-specific insights, and drill-down insights. Each insight is shown with a concise title and a list of key observations. These insights enable users to gain a deeper understanding of the current chart from different angles.


\section{Usage Scenario}
\label{sec: usage scenario}

\textcolor{black}{We illustrate the Intelligent Drill-Down through a scenario involving a public health analyst seeking to identify demographic subgroups potentially vulnerable to high stress. Using a  \href{https://www.kaggle.com/datasets/nagpalprabhavalkar/tech-use-and-stress-wellness}{ publicly available Kaggle stress wellness dataset } (25 variables) for illustrative purposes, the analyst employs our system to explore complex variable relationships that are difficult to uncover through manual analysis (Figure \ref{usage scenario}).}

\textcolor{black}{After uploading the dataset, the analyst queries:\textit{ “I want to analyze people whose stress level is above 7.”} The resulting visualization (Fig. \ref{usage scenario}a) reveals a noticeable concentration of highly stressed adolescents and suggests an association between shorter meditation duration and higher stress. To further examine regional patterns among younger individuals, the analyst uses the track interaction feature  (Fig. \ref{system}-c2) to filter the data to those under 28 and continues the drill-down with a focus on the relationship between region and stress level (Fig. \ref{usage scenario}-1), which indicates higher average stress among rural youth (Fig. \ref{usage scenario}b).}

\textcolor{black}{The analyst then backtracks (Fig. \ref{usage scenario}-2) using path navigation (Fig. \ref{system}b) and consults the insight panel (Fig. \ref{system}d), where the system provides automatically generated insight suggestions. Following a high-level recommendation, the analyst drills into the \textit{ High caffeine (\(\geq\) 180 mg/day)} dimension (Fig. \ref{usage scenario}-3). The resulting view (Fig. \ref{usage scenario}c) further reveals that, within the high-stress subgroup, higher caffeine intake is associated with poorer mood and reduced leisure time.}

\textcolor{black}{Overall, the system enables the analyst to efficiently characterize a potential high-risk profile and supports navigation across meaningful drill-down paths that might be overlooked in traditional workflows.}

\label{sec: User Study}
\section{User Study}
\textcolor{black}{Our primary goal was to evaluate whether the system enables users to discover more meaningful insights and select appropriate exploration paths compared to a baseline, while ensuring high usability. To this end, we conduct an end-to-end user study of the full workflow, using a controlled ablation baseline to isolate our incremental modules and counting only participant self-validated insights to conservatively control insight quality.}
\subsection{Experiment Setup}
\textbf{Study Conditions.} We adopted a within-group design in which each participant experienced two different experimental conditions.

\textbf{Baseline.} Users could only \textcolor{black}{perform drill-down operations} via natural language or basic recommended drill dimensions using our basic system. They were not permitted to record interactions or obtain system-generated insights or high-level recommended drill dimensions. 

\textbf{Intelligent Drill-Down.} Users could track their interactions and obtain insights and high-level recommended drill dimensions from the system. 

\textcolor{black}{\textbf{Controlled Ablation.} To isolate the incremental value of our proposed system-generated insights and drill-down suggestions grounded in the current exploration context, we used a controlled ablation baseline implemented within the same interface and infrastructure as the test condition. The baseline supports the same NL-driven chart specification and manual drill-down operations, but does not use interaction history for recommendation and does not present auto-generated insight candidates. Importantly, in both conditions participants were instructed to record only the insights they personally recognized and confirmed from the visual evidence; \textcolor{red}{system-generated texts (when available) served only as optional hypotheses and prompts. They were not counted unless the participant explicitly validated and wrote them down.} This design avoids UI/implementation confounds while enabling a fair comparison on whether adding system-generated insights and drill-down suggestions informed by the current exploration context improves path appropriateness and insight yield over a prevailing manual workflow. Regarding latency, both conditions shared the same LLM backend, ensuring uniform response times across trials.}

\textbf{Participants.} We invited 20 participants from social media (10 males and 10 females; 7 undergraduates and 13 postgraduates) to take part in the study through social media. They had not heard of Intelligent Drill-Down or seen the datasets involved.
Among these participants, 8 were majoring in computer science and technology and 12 in data science. All participants had data analysis experience. This selection ensured that they could complete data analysis tasks and evaluate their own performance.

\textbf{Datasets.}
We used two Kaggle datasets. One dataset examines the relationship between the use of technology in daily life and overall well-being (as in the \textcolor{red}{Sec. \ref{sec: usage scenario})}, while \href{https://www.kaggle.com/datasets/ziya07/carbon-trading-transactions-dataset}{the other} involves carbon trading records of publicly listed companies across multiple industries.
The former dataset was partially preprocessed. After processing, both datasets comprised 5,000 records and 15 columns. The datasets were equal in size and exhibited a consistent structure: each contained 8 numerical columns, 6 categorical columns, 1 temporal column, and no boolean or textual columns, indicating a uniform distribution of data types.
Both datasets provide a comprehensive set of aggregatable metrics and maintain balanced cardinalities across dimensions, making it well-suited for \textcolor{black}{tasks requiring deep drill-down exploration}.
For both datasets, we defined similar analytical tasks, and thus we considered them to be of comparable analytical complexity.



 
\textbf{Procedures. }
Before the experiment began, all participants were introduced to the concept of drill-down as well as the operation of the system. They were then given ten minutes to familiarize themselves with the system by working on a simple dataset.

Subsequently, participants engaged in data exploration under two distinct experimental conditions, each applied to a different dataset. The duration of each exploration session was fixed at 20 minutes. To mitigate order effects, both the sequence of experimental conditions and the assignment of datasets were counterbalanced across participants. Throughout the sessions, we encouraged participants to articulate insights. The study systematically recorded the total insights generated, the total nodes explored, and the corresponding time spent reasoning. Following each exploration task, participants were invited to quantitatively assess their performance across five dimensions: level of understanding of the dataset, appropriateness of the drill-down path, adequacy of the visualizations relative to expectations, degree of goal completion, and overall satisfaction.

After the experiment, user experience with the system was evaluated through questionnaires and semi-structured interviews.

\subsection{Experiment Result}

\begin{figure}[htbp]
    \centering
    \includegraphics[width=0.85\linewidth]{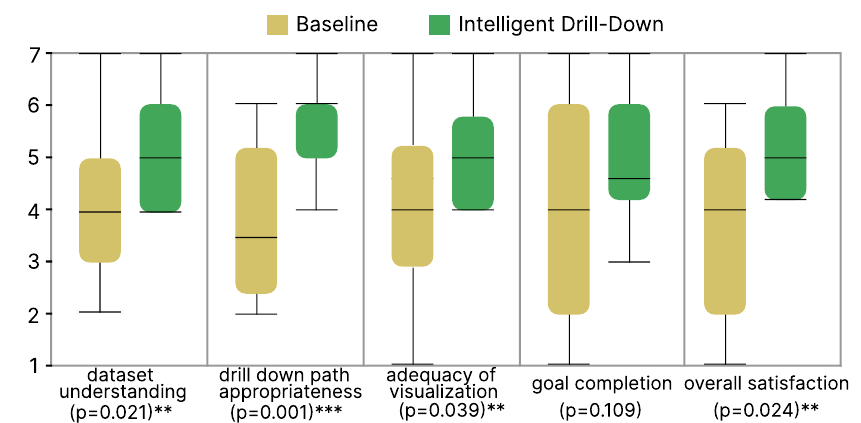}
    \caption{\textcolor{black}{Subjective ratings on a 7-point Likert scale (1 = Strongly Disagree, 7 = Strongly Agree).} The p value stands for significance level with *,**,*** stands for p<.1, .05 and .01, respectively.}
    \label{user study box}
\end{figure}

We employed the Wilcoxon signed-rank test to analyze the paired results. The significance levels were defined as follows: (*) for p < 0.1, (**) for p < 0.05, and (***) for p < 0.01. In the comparative experiment, we found that the number of insights identified when using Intelligent Drill-Down (6.6 ± 1.15) was, on average, two more than with the Baseline condition (3.9 ± 0.96), with a significant difference (p < 0.001***) (Fig. \ref{user study box}). 
\textcolor{red}{We compare the depth of insights users derived from the Baseline versus our system. Participants using the Baseline predominantly reported common-sense validation; for instance, in the Carbon Trading dataset, they identified \textit{a trivial negative correlation between trading volume and carbon price}. In contrast, users of our system uncovered counter-intuitive patterns, revealing that \textit{Certified cases surprisingly exhibit lower cost savings than Disputed ones}. A similar distinction exists in the Technology \& Health dataset: while the Baseline group reiterated known factors like \textit{mobile usage impairing sleep quality}, our system enabled users to perform variable decoupling. They discovered that \textit{working hours show no direct association with stress}, identifying instead that \textit{mindfulness practice is a critical mediator for low anxiety in females}. Ultimately, these findings demonstrate that our system empowers users to transcend surface-level validation, effectively facilitating the discovery of significantly higher-quality insights.} 
Furthermore, the significantly higher drill-down path appropriateness (p < 0.001) offers a proxy validation for our recommendation model (Sec. \ref{sec:drill-down path}), indicating effective prioritization of user-aligned paths.
With respect to thinking time, participants exhibited comparable performance across the two conditions. Interviews further revealed that, even with the aid of insights, users tended to carefully examine the data and consider their subsequent exploration steps. Regarding the number of nodes explored, Intelligent Drill-Down (4.0 ± 1.23) was slightly higher than the Baseline (3.0 ± 1.28), primarily due to differences in generation time. 
\textcolor{black}{Complementing the objective performance metrics above, to validate usability and perceived effectiveness, we compared participants' subjective evaluations (Fig.  \ref{user study box}).}
As shown, participants reported significantly higher satisfaction with the Intelligent Drill-Down system, particularly in terms of dataset comprehension, drill-down path selection \textcolor{black}{suggesting better intent alignment}, and overall user experience \textcolor{black}{reflecting generation quality}.

\begin{figure}[htbp]
    \centering
    \includegraphics[width=0.85\linewidth]{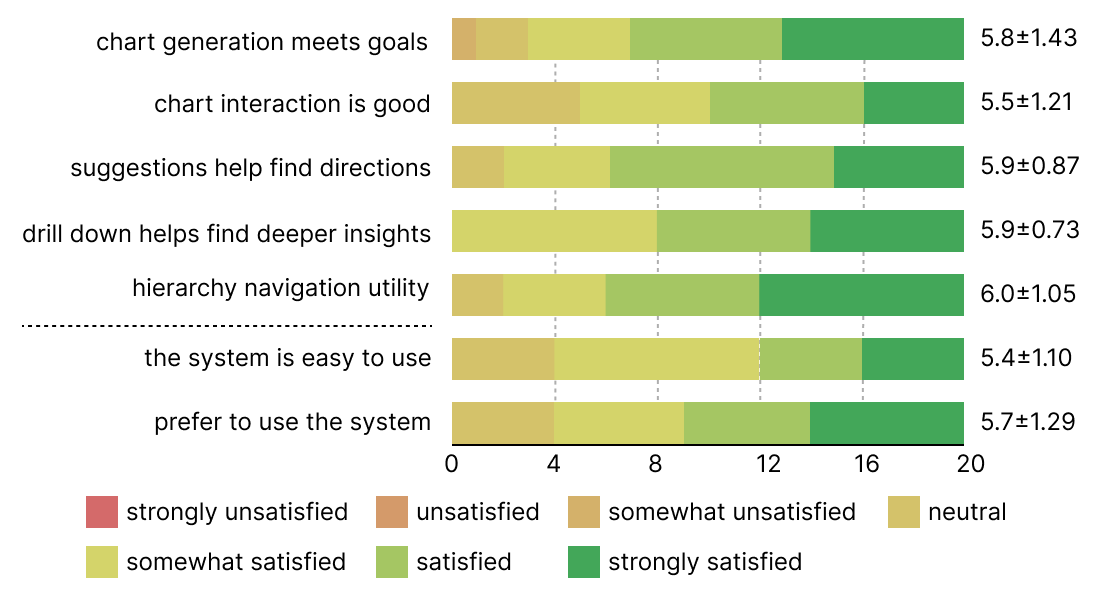}
    \caption{Results from the subjective questionnaires. The stack bars indicate feedback scores and the rightmost column shows Mean±STD.}
    \label{user study bar}
\end{figure}

The results of the questionnaire survey are shown in Figure \ref{user study bar}. The average user rating for our system was 5.65, with few participants expressing dissatisfaction with its various functions, suggesting a generally high level of user satisfaction. Feedback further indicated that users particularly appreciated the navigation bar feature, noting that it facilitated the discovery of deeper insights. Nevertheless, areas for improvement remain. In particular, users highlighted that the chart interaction was not sufficiently intelligent and that the system’s ease of use could be enhanced, as some participants found it somewhat challenging to get started.
Participants evaluated our system from the following perspectives.

\textbf{Recommended drill dimensions and generated insights}. Our system helps users identify drill-down paths. All participants utilized recommended drill dimensions and insight generation, and they believed these features assisted them in finding appropriate drill-down paths. In one case, a participant using the baseline system became overwhelmed by the multitude of available drill-down options and was unable to determine an optimal path forward. When faced with a similar scenario using our system, the participant selected the system-recommended drill dimensions and subsequently uncovered a significant amount of new analytical information. He told us,\textit{ The recommended drill dimensions helped me choose a suitable drill-down path, which greatly assisted my analysis process.} 

\textbf{Clear context management.} Our system’s clear context management supports users in conducting parallel drill-down explorations without becoming disoriented. In one case, a participant drew inspiration during the drill-down process of the second dimension and, using the parallel controller, fluidly switched back to the previous dimension without losing context; he then continued the drill-down and arrived at several new discoveries that built on his earlier findings. \textit{The parallel controller function is very practical. I can explore several paths in parallel and combine the information obtained from the drill-down.} he further told us, emphasizing how the parallel controller supported comparison, synthesis, and rapid iteration. 

\section{Discussion}






In this section, we mainly discuss the design trade-offs and limitations of our work and outline future directions.




\textbf{Relationship between interactions and user intent.}
In our work, we analyze user behavior based on recorded interactions. During the collection of interaction logs, noisy data may be introduced, which can interfere with the accurate inference of user intent. Our system grants users control over whether to capture interaction records, enabling them to determine which interactions reflect their analytical intent. This approach helps reduce noise introduced by interactions occurring during the exploratory phase. 
However, it is indeed possible for users to make errors, which may result in deviations in the outcomes, which was observed once during the user study in Sec. \ref{sec: User Study}.
\textcolor{black}{Additionally, our system excludes interactions shorter than \(500ms\) to ensure only meaningful and reliable data are collected~\cite{hansen2003command}.}
Several prior studies have proposed methods to address this issue, including categorizing user intent and recording it in a structured manner~\cite{ragan2015characterizing}, as well as mapping fine-grained interaction events to corresponding model updates~\cite{endert2012semantic}. In the future, we can leverage these methods to enhance the accuracy of LLMs in inferring user intent from interaction behavior via prompt engineering or fine-tuning.

\label{sec: Performance of LLMs}
\textcolor{black}{\textbf{Performance and Robustness of LLMs.} To validate the reliability and scalability of our prompt engineering akin to algorithmic verification, we conducted a quantitative evaluation focusing on two dimensions:
(1) Stability and Prompt Robustness: We evaluated the brittleness of our prompts by quantifying generation failures. With the prompt parameters fixed to minimize non-determinism (temperature=0.1, seed=42) ~\cite{fu2024hint,qin2024infobench}, the baseline model (GPT-5) exhibited a code execution error rate of \(9.7\%\) and a hallucination rate of 14\% in insight generation. While prompts demonstrated high stability across valid datasets, we implemented an automated rollback mechanism to handle the error cases. Future work may further reduce hallucinations by enhancing visual prompts over language priors ~\cite{favero2024multi}.
(2) Model Generalizability and Latency: To address concerns regarding reliance on proprietary models and potential obsolescence, we tested the system's cross-model validity. We adopted a parallelized processing approach \textcolor{red}{(avg. 45s end-to-end latency with GPT-5)}. While GPT-5 mini reduced latency by 54\%, it incurred a higher error rate (24.3\%). Crucially, we verified that the system functions with open-weights models like Qwen-Long. Although Qwen-Long currently lags in reasoning capability compared to GPT-5, its successful integration demonstrates that our framework is not strictly bound to a single proprietary provider and can adapt to locally deployed LLMs as open-source capabilities improve ~\cite{xu2024device}.}

\textcolor{black}{\textbf{Scalability and Generalizability.} We evaluated the system's applicability across data scale and domain diversity. (1) Data Scale: Due to browser rendering constraints, the system currently handles datasets up to 10 million cells. Future work will utilize hierarchical data cubes to extend this limit ~\cite{lins2013nanocubes}. (2) Domain Generalization: We validated the system across diverse domains, including environmental monitoring, public health, and education. By leveraging the LLM's inherent semantic knowledge and our domain-agnostic prompt engineering, the system effectively interprets data from these varied fields without requiring domain-specific fine-tuning.}



\textcolor{black}{\textbf{Balancing Approachability and User Agency.}
We address the potential tension between approachability and agency by strictly positioning our Intelligent Drill-down system as an intent translator not a decision replacement. While the system enhances approachability by proactively recommending insights and exploration paths, the user retains exclusive authority over the analytical direction. Whether steering through natural language, interaction tracking, or dimension clicking, the system’s role is to bridge the gulf of execution by converting high-level user intents into visual specifications without overriding user control. To further refine this equilibrium, future work could incorporate adaptive guidance to dynamically adjust automation levels by user expertise ~\cite{ceneda2016characterizing} and enhance algorithmic transparency to foster trust in automated suggestions ~\cite{shneiderman2020human}.}


\textcolor{red}{\textbf{Limitations of Insight Analysis and Visualization Design.}}
\textcolor{red}{To validate the system’s utility in facilitating knowledge discovery, our evaluation prioritized the volume of verified insights combined with a qualitative characterization of their nature. By ensuring a threshold of valid, high-quality findings, we demonstrated the system’s effectiveness in supporting user exploration~\cite{zhao2025proactiveva}. However, we acknowledge a limitation in the current analysis: we did not employ a standardized quantitative metric to evaluate the depth or complexity of individual insights. Future work will aim to develop more granular, data-driven metrics to provide a rigorous quantitative assessment of insight quality.}
\textcolor{black}{Currently, the system interface focuses on functional validation. In future work, we plan to refine the visualization system to improve the overall user experience.}
\section{Conclusion}

In conclusion, to address excessive path space, alignment difficulties with user intent, and disordered context management, we introduce Intelligent Drill-Down, an LLM-driven technique for human-AI collaborative visual exploration. Our framework leverages LLMs to generate visual insights and recommended drill dimensions, uses an interaction tracker to decompose user intent, and employs a hierarchy navigation component for context management. We implemented a system to demonstrate its effectiveness through a usage scenario and evaluation experiments.

\section*{Acknowledgments}
This work was supported by the Natural Science Foundation of China (NSFC No. 62472099).


\bibliographystyle{abbrv-doi}
\bibliography{template}

\end{document}